\PassOptionsToPackage{colorlinks}{hyperref}
\PassOptionsToPackage{allcolors=blue}{hyperref}
\documentclass[twocolumn,numberedappendix]{openjournal}
\pdfoutput=1
\usepackage{amsmath,amssymb,gensymb,color,graphicx,array,multirow,soul,rotating}
\usepackage{natbib,multirow,placeins,textcomp,subfigure}
\usepackage[outline]{contour}
\usepackage[utf8]{inputenc}
\usepackage[T1]{fontenc}
\usepackage{booktabs}
\usepackage{microtype}
\usepackage{orcidlink}
\usepackage{listings}
\usepackage{savesym}
\savesymbol{splitbox}
\usepackage[export]{adjustbox}

\newcommand{\vc}[1]{\begin{minipage}[c]{\linewidth}{\begin{center}#1\end{center}}\end{minipage}}

\newcolumntype{C}[1]{>{\centering\arraybackslash}m{#1}}

\DeclareFixedFont{\ttb}{T1}{txtt}{bx}{n}{9} 
\DeclareFixedFont{\ttm}{T1}{txtt}{m}{n}{9}  

\newcommand{\dfn}[1]{\textbf{#1}}

\newenvironment{closetabcols}[1][0.5mm]{\setlength{\tabcolsep}{#1}}{}

\newcommand\sinc{\,\text{sinc}}

\begin{document}

\title{Large-scale power loss in ground-based CMB mapmaking}

\author{Sigurd~Naess \orcidlink{0000-0002-4478-7111}}
\affil{Institute of Theoretical Astrophysics, University of Oslo, Norway}
\author{Thibaut~Louis \orcidlink{0000-0002-6849-4217}}
\affil{Université Paris-Saclay, CNRS/IN2P3, IJCLab, 91405 Orsay, France}


\begin{abstract}
CMB mapmaking relies on a data model to solve for the sky map, and this
process is vulnerable to bias if the data model cannot capture the full
behavior of the signal. We demonstrate that this bias is not just limited to
small-scale effects in high-contrast regions of the sky, but can manifest
as $\mathcal{O}(1)$ power loss on large scales in the map under conditions
and assumptions realistic for ground-based CMB telescopes. This bias is
invisible to simulation-based tests that do not explicitly model them,
making it easy to miss. We identify two different mechanisms that both
cause suppression of long-wavelength modes: sub-pixel errors and detector
gain calibration mismatch. We show that the specific case of subpixel bias can
be eliminated using bilinear pointing matrices, but also provide simple
methods for testing for the presence of large-scale model error bias in
general.
\vspace*{8mm}
\end{abstract}

\section{Introduction}

CMB telescopes observe the sky by scanning their detectors across
it while continuously reading off a series of samples from the
detectors. Typically the signal-to-noise ratio of each sample is
small, but by combining a large number of samples with knowledge
of which direction the telescope was pointing at any time, it's
possible to reconstruct an image of the sky. There are several
ways of doing this, with the most common being maximum-likelihood,
filter+bin and destriping. These all start by modeling the
telescope data as \citep{tegmark-mapmaking}
\begin{align}
	d = Pm + n \label{eq:model}
\end{align}
where $d$ is the set of samples read off from the detectors
(often called the time-ordered data), $m$ is the set of pixels
of the sky image we want to reconstruct, $n$ is the noise
in each sample (usually with significant correlations), and
$P$ is a pointing matrix that encodes how each sample responds
to the pixels in the image.

\begin{figure}
	\centering
	\begin{closetabcols}
	\begin{tabular}{cc}
		Input signal & Maximum likelihood solution \\
		\includegraphics[width=40mm]{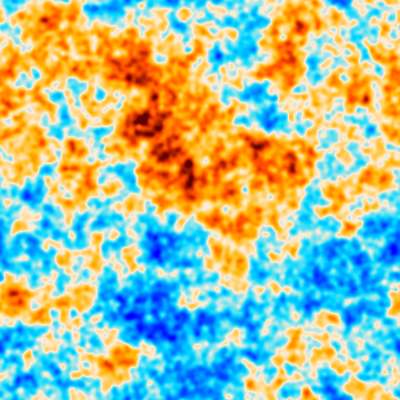} &
		\includegraphics[width=40mm]{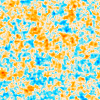} \\
	\end{tabular}
	\end{closetabcols}
	\caption{Preview of the model error bias we will discuss in the following
	sections. Despite the standard expectations that maximum-likelihood mapmaking
	is optimal and unbiased, the maximum-likelihood solution (\dfn{right})
	for a simple toy example is strongly power-deficient on large scales
	compared to the input signal (\dfn{left}). As we shall see this bias
	is not unique to maximum-likelihood methods, and can be triggered by
	several subtle types of model error.}
	\label{fig:maps-2d}
\end{figure}

Given this data model, it's possible to either directly solve for
an unbiased map (as in maximum likelihood mapmaking or destriping),
or to measure and correct for the bias in a biased estimator
(as in filter+bin mapmaking). In the case of maximum-likelihood
mapmaking, we solve for the least-squares solution
\begin{align}
	\hat m &= (P^TN^{-1}P)^{-1}P^TN^{-1}d
\end{align}
Linear regression like this is optimal and unbiased as long as only the
dependent variable ($d$ in this case) is noisy, but it is relatively
well known to be subject to \emph{attenuation bias}
if the independent variables ($P$ in this case)
are also noisy \citep{Draper1998-qf}. This is effectively what happens when we have model errors:
The $P$ we assume deviates from the true behavior of the instrument $P_\text{true}$
by some perturbation which effectively acts as a noise
source. Since $P$ enters into the denominator ($P^TN^{-1}P$) in the second power,
this noise term is squared, resulting in a noise bias there. We therefore end up
dividing by too much, leading to $m$ being underestimated.

Similar things happen in destriping and filter+bin mapmaking for much the same reason:
at some point, whether in the map-making or power spectrum estimation steps, one needs
to divide out the bias introduced by filtering, and if there's a noise bias in what's
being divided, then the result will be biased low.

In the context of CMB mapmaking, this bias has, to the extent that one has
been aware of it at all, been thought of as mainly a small-scale effect relevant only in
regions of the sky with very high contrast, leading to artifacts
like thin stripes or bowling around bright sources \citep{xgls-2017,model-error},
or as a tiny correction on intermediate scales (\citet{planck-ml-bias-2006}
saw a bias of 0.6\% peaking at $\ell=800$ for simulations of the Planck
space telescope). The goal of this paper is to
demonstrate the unintuitive and surprising result that that these effects
can lead to $\mathcal{O}(1)$ bias on large scales everywhere
in the map. The full scope of these errors appears to be largely
unappreciated, and we fear that some published results
may suffer from uncorrected bias at low multipoles in total intensity
due to these effects. As we shall see, it's mainly ground-based telescopes
that are vulnerable to this bias, which usually manifests as a power
deficit at large scales, as illustrated in figure~\ref{fig:maps-2d}.

In the following we will go through two common cases of model error in detail.
First subpixel errors, which arise when the smooth sky is approximated as
a mosaic of pixels; and then detector gain mismatch, which is becoming
increasingly important with today's $\mathcal{O}(10^5)$-detector telescopes.
We will demonstrate the effects with 1D and 2D toy examples and derive analytic
expressions for the bias; and test various mitigation methods such as reducing
the filtering of correlated noise.

\section{Subpixel errors}
\begin{figure*}
	\centering
	\begin{tabular}{cc}
		\dfn{\large a} & \dfn{\large b} \\
		\raisebox{-0.5\height}{\includegraphics[height=70mm]{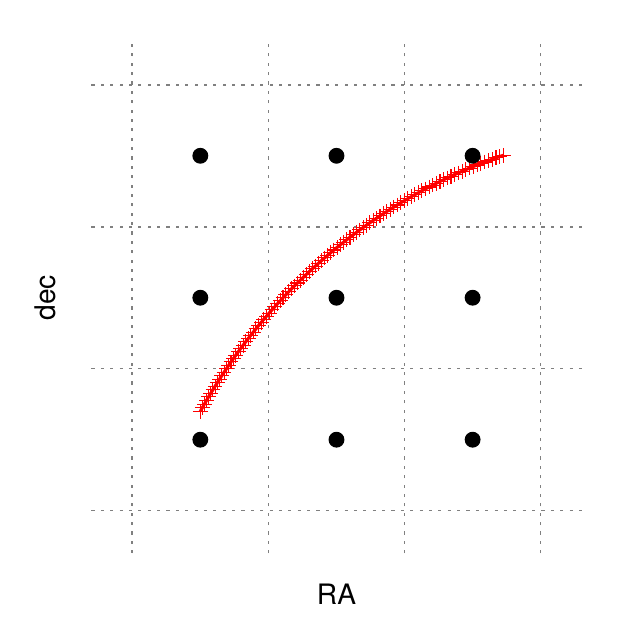}} &
		\hspace*{-5mm}\raisebox{-0.5\height}{\includegraphics[height=70mm]{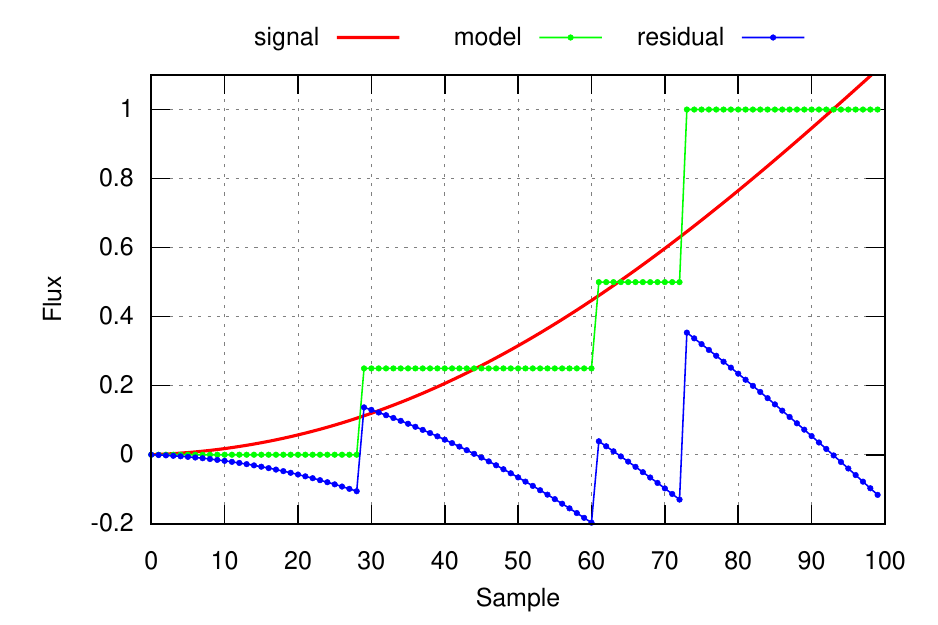}}
	\end{tabular}
	\caption{
		\dfn{a}: Example path (red) of a detector across a few pixels.
		The area closest to each pixel center (black dots) is shown with dotted lines.
		In the nearest neighbor model, the value associated with each sample is simply
		that of the closest pixel, regardless of where inside that pixel it is.
		\dfn{b}: Example detector signal (red) for the same path. The closest
		matching model (green) leaves a jagged residual (blue) that has power on
		all length scales despite the signal itself being very smooth.
		For comparison, if our model were a constant zero, then the residual
		would just be the signal itself (red), and hence smooth.
		{\bf If smooth residuals are much cheaper in the likelihood than jagged ones,
		then a zero model will be preferred to one that hugs the signal as
		tightly as possible} like the green curve.
	}
	\label{fig:nearest-neigh}
\end{figure*}
Subpixel errors may be both the most common and most important
class of model errors in CMB mapmaking.
For efficiency reasons $P$ is almost always\footnote{
	While methods with non-discontinuous pointing matrices exist, e.g.
	\citet{artdeco}, these have had very limited adoption. As far as we
	are aware, every CMB telescope project has used nearest neighbor
	mapmaking for their official maps. This includes destripers like MADAM
	\citep{madam/2010} and SRoll \citep{planck/hfi/maps/2020} used in Planck,
the filter+bin map-makers used in SPT \citep{spt/maps/2011,spt/2021} and BICEP
\citep{bicep2a/2014} and the maximum likelihood map-makers used in ACT \citep{aiola/2020}
and QUIET \citep{quiet-gal/2015}.
}
chosen to use simple nearest-neighbor interpolation, where the
value of a sample is simply given by the value of the pixel nearest
to it. This means that $P$ can be implemented by simply reading off
one pixel value per sample, and its transpose $P^T$ consists of simply
summing the value of the samples that hit each pixel. However, this
comes at the cost of there being a discontinuous jump in values as one
scans from one pixel to the next, as illustrated in figure~\ref{fig:nearest-neigh}.
Hence, the closest the model can get to a smooth curve is a
staircase-like function that hugs it, leaving a residual full of
discontinuous jumps (the blue curve in the figure).

\begin{figure}
	\centering
	\hspace*{-2mm}\includegraphics[width=1.05\columnwidth]{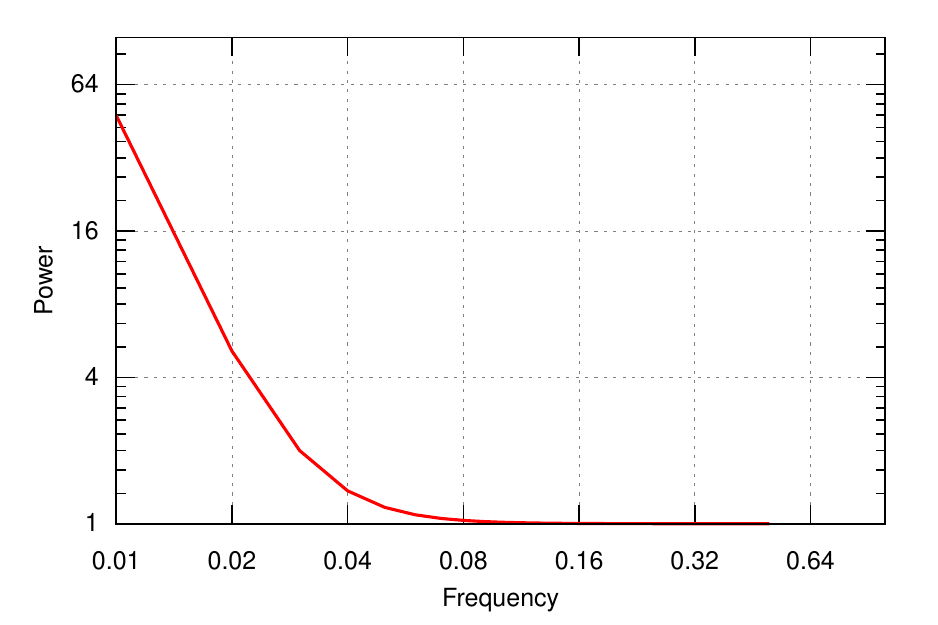}
	\caption{
		The noise model/inverse weights/inverse filter used in the subpixel
		bias demonstration in figures~\ref{fig:subpix-bias} and \ref{fig:subpix-noerr}.
		It is a simple Fourier-diagonal 1/f + white noise spectrum
		typical for ground-based CMB observations. The frequency axis is in
		dimensionless units in this toy example, but for real telescopes the
		transition from white noise is typically a few Hz, corresponding
		to multipoles of hundreds on the sky. The power axis is dimensionless
		here, but for a real-world case could have units like $\micro$K$^2$/Hz.
	}
	\vspace{3mm}
	\label{fig:ps}
\end{figure}

Discontinuous residuals are not necessarily problematic. The trouble arises
when this is coupled with a likelihood\footnote{The equivalent for filter+bin is a filter that
impacts some modes more than others (which is the whole point of a filter),
and for destriping it's the baseline length and any amplitude priors.
} where some modes have much less weight than others. Ground-based CMB
telescopes suffer from atmospheric emission that acts as a large source
of noise on long timescales. This leads to time-ordered data noise power
spectra similar to the one sketched in figure~\ref{fig:ps}, with
a white noise floor at short timescales (high frequency) transitioning to a steep
rise of several orders of magnitude as one moves to longer timescales
(low frequency). In this case long-wavelength modes have orders of magnitude
lower weight in the likelihood than short-wavelength modes. Put another way,
they are orders of magnitude \emph{cheaper to sacrifice} when the model can't fully
fit the data.

\subsection{1D toy example}
To illustrate the interaction between a nearest neighbor pointing matrix's
subpixel errors and a likelihood where large scales have low weight, let
us consider a simple 1D case where 100 samples scan uniformly across 10 pixels\footnote{
A real survey might have $\sim 3$ samples per pixel per scan for a single detector,
but would end up with much more than that across multiple scans and multiple detectors.}:
\begin{lstlisting}
npix  = 10
nsamp = 100
pix   = np.arange(nsamp).astype(float)*npix/nsamp
\end{lstlisting}
A standard nearest-neighbor pointing matrix for this looks like:
\begin{lstlisting}
P = np.zeros((nsamp,npix))
for i, p in enumerate(pix):
	P[i,int(np.round(pix[i]))%npix] = 1
\end{lstlisting}
We assume a typical Fourier-diagonal ``1/f'' noise model
$N(f) = 1+(f/f_\text{knee})^\alpha$
with $f_\text{knee}=0.03$ and $\alpha=-3.5$, and build the
inverse noise matrix/filter/baseline-prior $F$ by
projecting the inverse noise spectrum $N^{-1}$ into pixel space:\footnote{
	In practice the ``1/f'' power law does not continue to infinity at zero
	frequency, but stabilizes at some high value at low frequency. We implement
	this by replacing $f=0$ with a small, non-zero number for the purposes of
	computing the noise spectrum in the following.
}
\begin{lstlisting}
freq   = np.fft.rfftfreq(nsamp)
inv_ps = 1/(1+(np.maximum(freq,freq[1]/2)/0.03)**-3.5)
F = np.zeros((nsamp,nsamp))
I = np.eye(nsamp)
for i in range(nsamp):
	F[:,i] = np.fft.irfft(inv_ps*np.fft.rfft(I[i]), n=nsamp)
\end{lstlisting}
The signal itself consists of just a long-wavelength sine wave:\footnote{
Since all the mapmaking methods we will discuss are linear,
signal and noise can analysed independently. Since the focus
in this example is bias, it's therefore enough to consider a
signal-only data set, and we thus don't add any noise.}
\begin{lstlisting}
signal = np.sin(2*np.pi*pix/npix)
\end{lstlisting}

With this in place, we can now define our map estimators.

\subsubsection{Binning}
The \emph{binned map} is simply the mean value of the samples
in each pixel, with no weighting:
\begin{lstlisting}
map_binned = np.linalg.solve((P.T.dot(P)), P.T.dot(signal))
\end{lstlisting}
This is the unweighted least-squares solution for the map.
With a nearest neighbor pointing matrix where each sample
only affects one pixel $P^TP$ is diagonal, making solving
for the binned map very fast.

\subsubsection{Maximum-likelihood}
The maximum-likelihood solution of equation~\ref{eq:model} for the
sky image $m$ is
\begin{align}
	\hat m = (P^TN^{-1}P)^{-1}P^TN^{-1}d
\end{align}
where $N$ is the covariance matrix of the noise $n$. This is the
generalized least-squares solution for the map. Unliked binned
mapmaking, the matrix $P ^TN^{-1}P$ except for the special case
of uncorrelated noise. Solving for the maximum-likelihood map
can therefore be hundreds of times slower, and requires
iterative methods for large maps.\footnote{
	The matrix $P^TN^{-1}P$ is often poorly conditioned, making it slow
	for iterative solution methods to recover the large scales. If iteration is
	stopped early, this could result in a lack of power at large scales.
	This well-known phenomenon is \emph{not} the bias we're exploring in
	this paper, and since we solve the equation exactly it does not
	appear here.
}
In our toy example, $N^{-1} = F$, so the \emph{maximum-likelihood map}
is
\begin{lstlisting}
map_ml = np.linalg.solve((P.T.dot(F).dot(P)),P.T.dot(F.dot(signal)))
\end{lstlisting}

\subsubsection{Filter+bin}
As the name suggests, filter+bin consists of filtering the time-ordered
data, and then making a binned map. We'll use $F$ as our filter, so
the \emph{filter+bin map} is
\begin{lstlisting}
map_fb = np.linalg.solve(P.T.dot(P), P.T.dot(F).dot(signal))
\end{lstlisting}
The filter+bin map is biased by design, so to interpret or debias it one
needs to characterize this bias. There are two common approaches
to doing this: Observation matrix and simulations.

\paragraph{Observation matrix}
The observation matrix approach
recognizes that the whole chain of operations observe, filter, map
together make up a linear system, and can therefore be represented
as a matrix, called the \emph{observation matrix} \citep{bicep2-obsmat}.
Building this matrix is heavy, but doable for some
surveys. Under the standard assumption that the observation step is
given by equation~\ref{eq:model}, the observation matrix is given by
\begin{lstlisting}
obsmat = np.linalg.inv(P.T.dot(P)).dot(P.T.dot(F).dot(P))
\end{lstlisting}
and using it, we can define a debiased filter+bin map
\begin{lstlisting}
map_fb_deobs = np.linalg.solve(obsmat, map_fb)
\end{lstlisting}

\paragraph{Simulations}
Alternatively, and more commonly, one can characterize the bias
by simulating the observation of a set of random skies, passing
them through the filter+bin process, and comparing the properties
of the input and output images \citep[e.g.][]{spt-bmodes-2020}. The standard way of doing this is
by assuming that equation~\ref{eq:model} describes the observation
process, and that the bias can be described by a \emph{transfer function}:
a simple independent scaling of each Fourier mode. Under
these assumptions, we can measure and correct the bias as follows.
\begin{lstlisting}
nsim = 1000
sim_ips = np.zeros(npix//2+1)
sim_ops = np.zeros(npix//2+1)
for i in range(nsim):
	sim_imap = np.random.standard_normal(npix)
	sim_omap = np.linalg.solve(P.T.dot(P), P.T.dot(F).dot(P).dot(sim_imap))
	sim_ips += np.abs(np.fft.rfft(sim_imap))**2
	sim_ops += np.abs(np.fft.rfft(sim_omap))**2
tf = (sim_ops/sim_ips)**0.5
map_fb_detrans = np.fft.irfft(np.fft.rfft(map_fb)/tf, n=npix)
\end{lstlisting}

\subsubsection{Destriping}
Destriping splits the noise into a correlated and uncorrelated part,
and models the correlated noise as a series of slowly changing
degrees of freedom to be solved for jointly with the sky image itself
\citep{descart-destriper,planck-destriping}.
The data is modeled as
\begin{align}
	d &= Pm + Qa + n_w
\end{align}
where $n_w$ is the white noise with diagonal covariance matrix $N_w$,
and $Q$ describes how each correlated noise degree of freedom $a$
maps onto the time-ordered data, typically in the form of seconds
(ground) to minutes (space) long baselines. Given this, the
maximum-likelihood solutions for $a$ and $m$ are
\begin{align}
	Z &= I - P(P^TN_w^{-1}P)^{-1}P^TN_w^{-1} \notag \\
	a &= (Q^TN_w^{-1}ZQ + C_a^{-1})^{-1} Q^TN_w^{-1}Zd \notag \\
	m &= (P^TN_w^{-1}P)^{-1}P^TN_w^{-1}(d - Qa) \label{eq:destripe}
\end{align}
where $C_a$ is one's prior knowledge of the covariance of $a$.
Similarly to binned and filter+bin mapmaking, destriping derives its speed
from the diagonality of $P^TN_w^{-1}P$ allowing for instant inversion.
On the other hand, the equation for the baseline amplitudes $a$ is
not diagonal, and destriping allows a speed/optimality tradeoff in the choice of
baseline length and prior. It approaches the maximum likelihood
map when the baseline length is a single sample and $C_a + N_w = N$.
We implement this limit below, but explore other choices in section~\ref{sec:toy-2d}.
In our toy example $N_w = I$, so $C_a = F^{-1}-I$.
\begin{lstlisting}
iCa = np.linalg.inv(np.linalg.inv(F) - I)
Z   = I-P.dot(np.linalg.solve(P.T.dot(P), P.T))
a   = np.linalg.solve(Z+iCa, Z.dot(signal))
map_ds = np.linalg.solve(P.T.dot(P), P.T.dot(signal - a))
\end{lstlisting}

\subsubsection{Results}
\begin{figure}
	\centering
	\hspace*{-5mm}\includegraphics[width=1.07\columnwidth]{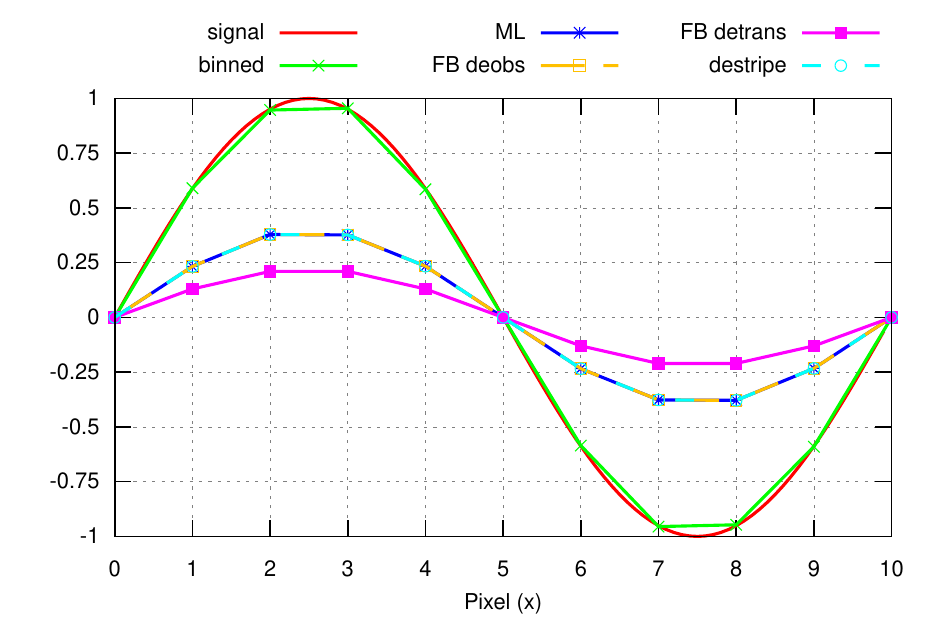}
	\caption{
		Demonstration of large loss of power in long-wavelength mode
		caused by the poor subpixel treatment in the standard nearest-neighbor pointing matrix.
		The vertical axis is dimensionless in this toy example, but
		could have units like $\micro$K or Jy/sr for a real-world case.
		Figure~\ref{fig:ps} shows the noise model/inverse weights/inverse filter
		used in the various methods.
		\dfn{signal}: The input signal, a smooth long-wavelength mode,
		sampled at 10 samples per output pixel.
		\dfn{binned}: Simple binned map (the unweighted average per pixel).
		Very suboptimal in the presence of correlated noise, but unbiased.
		\dfn{ML}: Maximum-likelihood map. 2/3 of the signal amplitude is lost despite
		the naive expectation of biaslessness for this estimator.
		\dfn{FB deobs}: Filter+bin map debiased using an observation matrix.
		Identical to ML.
		\dfn{FB detrans}: Filter+bin map debiased by deconvolving a
		transfer function measured from simulations. Even more biased
		than the others due to ignoring mode coupling.
		\dfn{destripe}: Destriper in the maximum-likelihood limit
		(1-sample baselines with optimal baseline prior). Identical to ML.
	}
	\vspace{3mm}
	\label{fig:subpix-bias}
\end{figure}

Figure~\ref{fig:subpix-bias} compares the recovered 1D sky ``images''
for the different mapmaking methods for this toy example. All methods
are expected to have a small loss of power at small scale (called
the ``pixel window'') due to averaging the signal within each pixel,
but this effect is well-known, easy to model, and not our focus here.
We deconvolve it using the following function before plotting.
\begin{lstlisting}
def dewin(x): return np.fft.irfft(np.fft.rfft(x)/np.sinc(freq),n=len(x)).real
\end{lstlisting}
After pixel window deconvolution the binned map (green) closely matches the
input signal (red). The same can not be said for the other estimators.
Maximum likelihood, filter+bin with observation matrix debiasing and destriping
(which are all equivalent in the limit we consider here) are strongly biased,
with the signal only being recovered with 1/3 of its real amplitude.

The situation is even worse for filter+bin with simulation-based debiasing,
as this suffers from an additional bias due to assuming that the Fourier modes
are independent.\footnote{This additional bias disappears if the simulations
have exactly the same statistical properties as the real signal we wish to
recover.}

\subsubsection{Explanation}
To see why inaccuracies in modeling the signal at sub-pixel scales can bias
the largest scales in the map, let us consider the example in figure~\ref{fig:nearest-neigh},
where a detector measures a smooth, large-scale signal while moving across a few
pixels. With a nearest-neighbor pointing matrix it is impossible to model this
smooth signal: the model for each sample is simply that of the closest pixel,
regardless of where inside that pixel it is. The model therefore looks like
a staircase-like function in the time domain.

Given that we can't exactly match the signal, what is the best approximation?
Let us consider two very different alternatives. \dfn{Model A}: The value in each pixel
is the average of the samples that hit it, making the model curve trace the
smooth signal as closely as it can. This is the green curve in figure~\ref{fig:nearest-neigh},
and has the sawtooth-like residual shown with the blue curve. It is probably the
model most people would choose if asked to draw one manually.
\dfn{Model B}: The value in each pixel is zero, and the residual is simply the signal itself.
Model B seems like a terrible fit to the data, but under a reasonable noise model
for a ground-based telescope, like the one shown in figure~\ref{fig:ps}, it
will actually have a higher likelihood (lower $\chi^2 = r^TN^{-1}r$
where $r$ is the residual) than model A. The reason is that while model B has a much
larger residual than model A, model B's residual is smooth and hence has most of its
power at low frequency in the time domain where $N^{-1}$ is very small. Meanwhile, model A's
residual extends to all frequencies due to its jagged nature, including the costly high
frequencies. The actual maximum-likelihood solution will be intermediate between these
two cases, sacrificing some but not all of the large-scale power in the model to make
the residual smoother.

\begin{figure}
	\centering
	\hspace*{-5mm}\includegraphics[width=1.07\columnwidth]{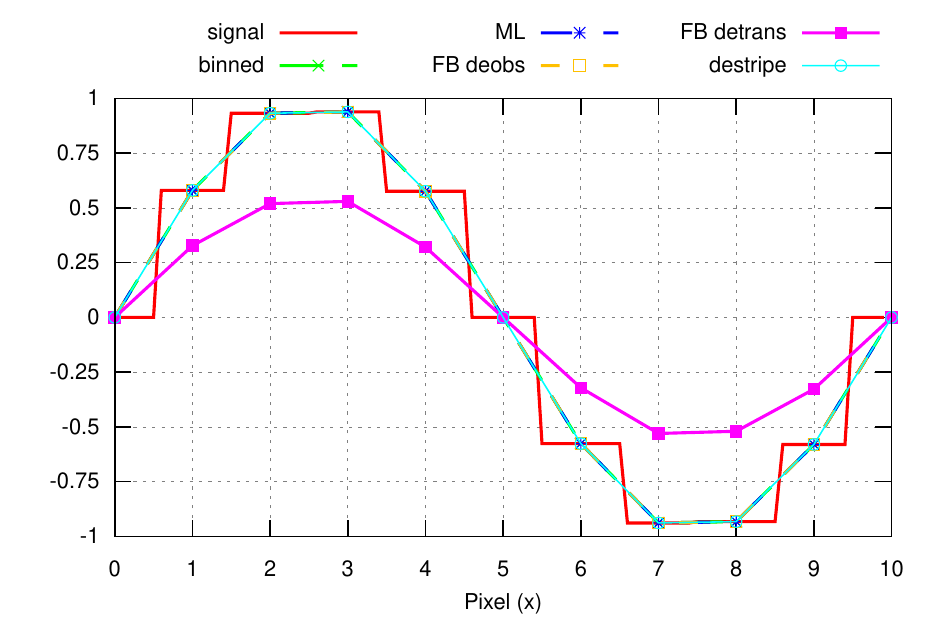}
	\caption{
		Like figure~\ref{fig:subpix-bias}, but with the input signal
		having the same nearest-neighbor pixelization as the models.
		In this case all models except FB detrans are unbiased.
	}
	\label{fig:subpix-noerr}
\end{figure}

To demonstrate that the bias really is caused by subpixel errors,
we repeated the simulation with a signal that follows the same nearest-neighbor
behavior as the data model, thus eliminating subpixel errors. The
result is shown in figure~\ref{fig:subpix-noerr}. The bias has disappeared
in all methods except filter+bin with transfer-function based debiasing,
which has an additional source of bias due to its assumption of a Fourier-diagonal
filter response.

\subsection{2D toy example}
\label{sec:toy-2d}
The 1D toy example is useful for understanding the origin of the bias, but
its unrealistic observing pattern makes it insufficient for exploring
optimality/bias tradeoffs in the different methods. We therefore made a larger
toy example where a single detector scans at constant speed across a square
patch, sampling $N_\text{scan} = 400$ equi-spaced rows with $N_\text{scan}$
equi-spaced samples per row, followed by a column-wise scan the same patch,
leading to a total cross-linked data set $N_\text{samp} = 2 N_\text{scan}^2 = 3200$
samples long. This equi-spaced grid of samples was chosen to make it easy to
simulate a signal directly onto the samples without needing the complication
of pixel-to-sample projection. These samples will then be mapped onto a
square grid of pixels with a side length of $N_\text{side} = 100$ using
the different mapmaking methods.

For the signal we draw realizations from a CMB-like $1/l^2$ power spectrum
with a Gaussian beam with standard deviation $\sigma=3$ pixels. Here $l$
is the pixel-space wave-number, which we evaluate in the higher
resolution sample space as
\begin{lstlisting}
ly = np.fft.fftfreq(N_scan)[:,None] * N_side/N_scan
lx = np.fft.fftfreq(N_scan)[None,:] * N_side/N_scan
l  = (ly**2+lx**2)**0.5
\end{lstlisting}
With this we can define the signal power spectrum $C_l = l^{-2}$ and beam $B_l = \exp(-l^2 \sigma^2/2)$,
and draw signal realizations as
\begin{lstlisting}
signal_map = np.fft.ifft2(np.fft.fft2(np.random.standard_normal((N_scan,N_scan)))*Cl**0.5*Bl).real
signal_tod = np.concatenate([signal_map.reshape(-1), signal_map.T.reshape(-1)])
\end{lstlisting}
The last step here takes into account that the simulated scanning pattern covers the
field twice, first horizontally and then vertically.

For the noise we use a simple 1/f spectum $N(f) = 1 + (f/f_\text{knee})^\alpha$,
with \lstinline{f = np.fft.rfftfreq(N_samp)}
and the knee frequency $f_\text{knee}$ corresponding to 1/30th of the side length,
$f_\text{knee} = 0.5\cdot 30/N_\text{scan} = 0.0375$ (6.7 pixel wavelength), and with
an atmosphere-like exponent $\alpha=-3.5$.

The pixel coordinates of each sample are
\begin{lstlisting}
pix_pat1 = (np.mgrid[:N_scan,:N_scan]*N_side/N_scan).reshape(2,-1)
pix_pat2 = pix_pat1[::-1]
pix = np.concatenate([pix_pat1,pix_pat2],1)
\end{lstlisting}
which we use to build the nearest-neighbor pointing matrix
\begin{lstlisting}
iy, ix = np.floor(pix).astype(int)%N_side
P_nn   = scipy.sparse.csr_array((np.full(N_samp,1),(np.arange(N_samp),iy*N_side+ix)),shape=(N_samp,N_pix))
\end{lstlisting}

\begin{figure*}[p]
	\centering
	\hspace*{-10mm}\begin{closetabcols}
		\begin{tabular}{C{13mm}ccm{1mm}|m{1mm}ccC{14mm}}
			& Signal & Noise & & & Noise & Signal & \\
		Input &
		\includegraphics[width=30mm,valign=m]{subpix/toy2d_input_signal_map.png} &
		& & &
		\includegraphics[width=30mm,valign=m]{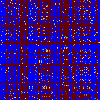} &
		\includegraphics[width=30mm,valign=m]{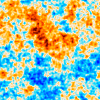} &
		bin \\[13.6mm]
		ML &
		\includegraphics[width=30mm,valign=m]{subpix/toy2d_ml_nn_signal_map.png} &
		\includegraphics[width=30mm,valign=m]{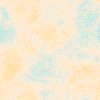} & & &
		\includegraphics[width=30mm,valign=m]{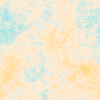} &
		\includegraphics[width=30mm,valign=m]{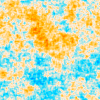} &
		\vc{DS 4}\\[13.6mm]
		\vc{ML w1}&
		\includegraphics[width=30mm,valign=m]{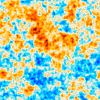} &
		\includegraphics[width=30mm,valign=m]{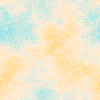} & & &
		\includegraphics[width=30mm,valign=m]{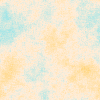} &
		\includegraphics[width=30mm,valign=m]{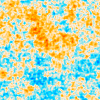} &
		\vc{DS+ 4}\\[13.6mm]
		\vc{ML w2}&
		\includegraphics[width=30mm,valign=m]{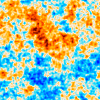} &
		\includegraphics[width=30mm,valign=m]{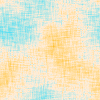} & & &
		\includegraphics[width=30mm,valign=m]{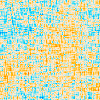} &
		\includegraphics[width=30mm,valign=m]{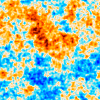} &
		\vc{DS 64}\\[13.6mm]
		\vc{ML w3}&
		\includegraphics[width=30mm,valign=m]{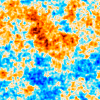} &
		\includegraphics[width=30mm,valign=m]{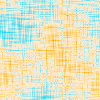} & & &
		\includegraphics[width=30mm,valign=m]{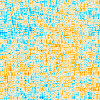} &
		\includegraphics[width=30mm,valign=m]{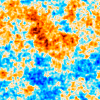} &
		\vc{DS+ 64}\\[13.6mm]
		\vc{ML lin} &
		\includegraphics[width=30mm,valign=m]{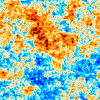} &
		\includegraphics[width=30mm,valign=m]{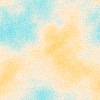} & & &
		\includegraphics[width=30mm,valign=m]{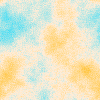} &
		\includegraphics[width=30mm,valign=m]{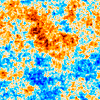} &
		\vc{DS+ 4 lin}
	\end{tabular}
	\end{closetabcols}
	\caption{
		Example signal and noise maps for the different mapmaking methods.
		The top left map is the input signal, which is directly evaluated
		at 4x higher resolution than what is used for reconstructing the output maps.
		All output maps were built from the same data and assume a nearest neighbor pointing matrix,
		except for the last row where a bilinear pointing matrix was used.
		The color range is the same for all panels.
		The binned map (top right) is bias-free but has uselessly high noise. Maximum likelihood (ML)
		and destriping with short baselengths (with (DS+) and without (DS) baseline amplitude prior,
		which only matters for the shortest baselines)
		are all low-noise but biased on large scales. This bias goes away as the noise model
		is artificially whitened (ML w$X$) or the baseline length is increased (for destriping),
		but this comes at a cost of increased noise on all scales.
		Bilinear mapmaking (last row) instead avoids the bias by eliminating most of the model error.
		It can be difficult to judge how significant the bias and noise levels are from
		these images. See figures~\ref{fig:2d-bias} and \ref{fig:2d-noise} for easier to
		interpret comparisons of the signal and noise power spectra respectively.
	}
	\label{fig:2d-maps}
\end{figure*}

\begin{figure}
	\centering
	\hspace*{-5mm}\includegraphics[width=1.10\columnwidth]{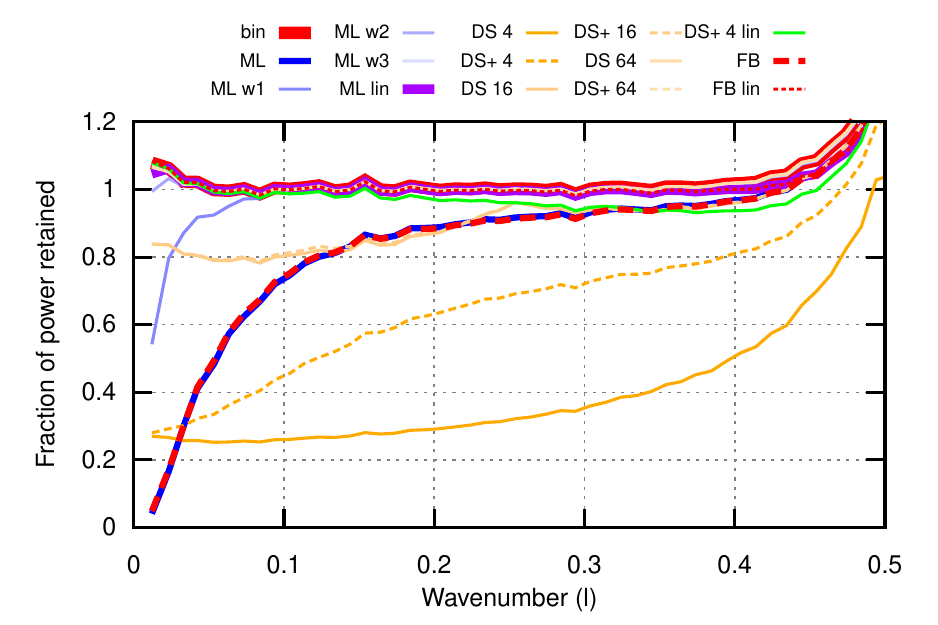}
	\caption{
		Comparison of the subpixel bias of different mapmaking methods
		described in section~\ref{sec:2d-cases} and discussed in section~\ref{sec:2d-results}.
		Standard maximum-likelihood (thick blue) is strongly biased, as are all but
		the (very noisy) longest destriping baseline. This bias disappears (ML)
		or is greatly reduced (DS) when switching to bilinear interpolation in the pointing matrix.
		See figure~\ref{fig:2d-noise} for the corresponding noise spectra.
		The wavenumber $l$ is dimensionless and with a Nyquist frequency of 0.5.
		we interpret the upturn at $l>0.4$ as aliasing, which
		is expected and not relevant for the biases we consider here.
	}
	\label{fig:2d-bias}
\end{figure}

\begin{figure}[h!]
	\centering
	\hspace*{-5mm}\includegraphics[width=1.10\columnwidth]{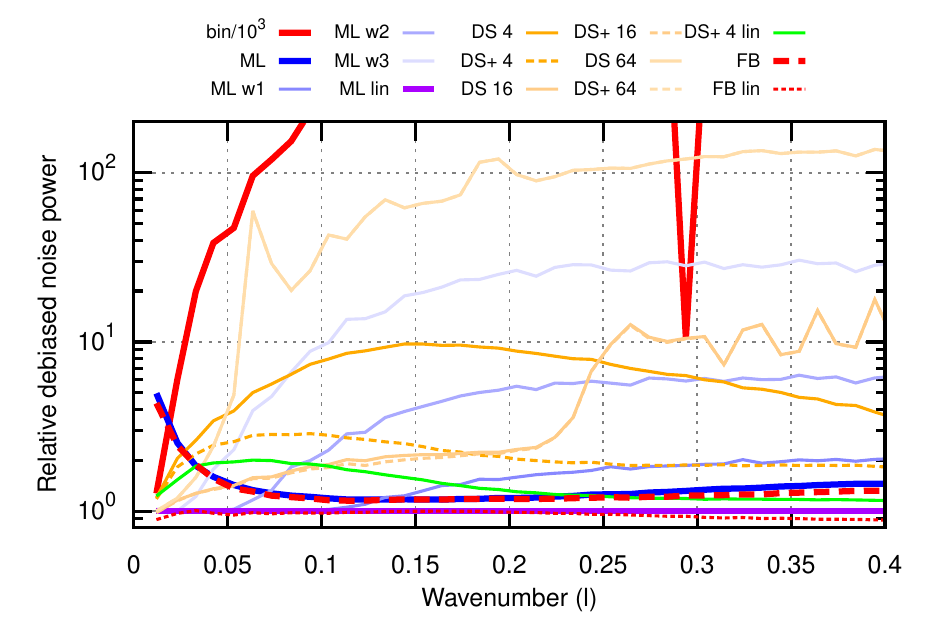}
	\caption{
		Comparison of the optimality for the methods defined in section~\ref{sec:2d-cases}, measured
		as the debiased noise power spectrum (using the bias measurements from figure~\ref{fig:2d-bias})
		for each method relative to that of bilinear maximum-likelihood mapmaking.
		Note the logarithmic vertical axis, and that the binned case was divided by $10^3$
		to bring it partially inside the plot bounds.
		The wavenumber $l$ is dimensionless and with a Nyquist frequency of 0.5.
		It's clear that simple bias mitigation
		methods like artificially whitening the noise model or increasing the
		baseline length are too costly to be practical.
	}
	\label{fig:2d-noise}
\end{figure}

\subsubsection{Cases}
\label{sec:2d-cases}
We will investigate 5 classes of nearest-neighbor maps:
\begin{enumerate}
	\item \dfn{bin}: Simple binned map, which we expect to be unbiased but very noisy
	\item \dfn{ML}: Maximum-likelihood map. Ideally unbiased and optimal, but will deviate
		from this due to model errors.
	\item \dfn{ML w$X$}: Maximum-likelihood maps with a ``whitened'' noise model,
		which overestimates the
		white noise power by a factor $10^X$, $X\in\{1,2,3\}$, reducing the overall
		correlatedness of the noise model. We expect the bias to be proportional to the
		overall dynamic range of the noise model,
		so making the noise model whiter should reduce bias. The cost will be suboptimal noise
		weighting, leading to a noisier map, but it might be worth it.
	\item \dfn{DS $X$}: Destriping map with a baseline of $X\in \{4,16,64\}$ samples and no amplitude
		prior. This is probably the most common type of destriping. The baseline lengths
		can be compared with the noise knee wavelength of 6.7 pixels $\approx$ 27 samples.
		This is the wavelength where the correlated and white noise powers are equal.
	\item \dfn{DS+ $X$}: Like DS $X$, but uses the correlated part of the maximum-likelihood
		noise model as an amplitude prior ($C_a$ in equation~\ref{eq:destripe}).
	\item \dfn{FB}: Filter+bin mapmaking where a transfer function is measured using
		simulations based on the data model, and this is deconvolved when measuring
		the power spectrum. This is the standard approach for filter+bin mapmaking.
		In this case only the power spectrum is debiased, not the map, so we don't
		show an example map for this case.
\end{enumerate}

Since all the mapmaking methods are linear, the signal and noise can be mapped separately.
We make 400 signal-only and noise-only data realizations, and map them using each method,
computing the mean signal and noise power spectra.

\subsubsection{Results}
\label{sec:2d-results}
Example maps are shown in figure~\ref{fig:2d-maps},
while the bias and noise are quantified in figures~\ref{fig:2d-bias} and \ref{fig:2d-noise}
respectively.
As expected the binned map is unbiased but extremely noisy. The maximum-likelihood
map is low-noise, but measurably biased on all scales, with a power deficit of
a few percent at the smallest scales which grows to almost 100\% on the largest scales.
The deficit appears to fall proportionally with the whitening, with ML w1 and w2
being respectively 10x and 100x as close to unbiased. Sadly this comes at the cost of
40\% and 350\% higher noise power respectively. These numbers may differ for real-world
cases, but this still seems like a very expensive bias mitigation method.
All but the longest-baseline destriped maps are also strongly biased,
with the shortest baseline being considerably worse than ML for almost all scales.
Much like we saw with the ML variants, the less biased destriping versions
come at a high cost in noise. Finally, the filter+bin map is biased even after simulation-based
debiasing due to the simulations not capturing the subpixel behavior of the real data.
Both the bias and noise levels are the same as maximum-likelihood in this example.

To test whether the observed biases are truly caused by subpixel errors, we repeat
the simulations with only one sample per pixel ($N_\text{scan} = N_\text{side}$).
As expected this results in an unbiased power spectrum for all methods.\footnote{
	Even filter+bin, which ignores mode coupling, ends up having an unbiased power
	spectrum because the simulations used to build the transfer function followed
	the same distribution as the real signal.
}

\subsubsection{Effective mitigation of subpixel errors}
\begin{figure}
	\centering
	\includegraphics[width=\columnwidth]{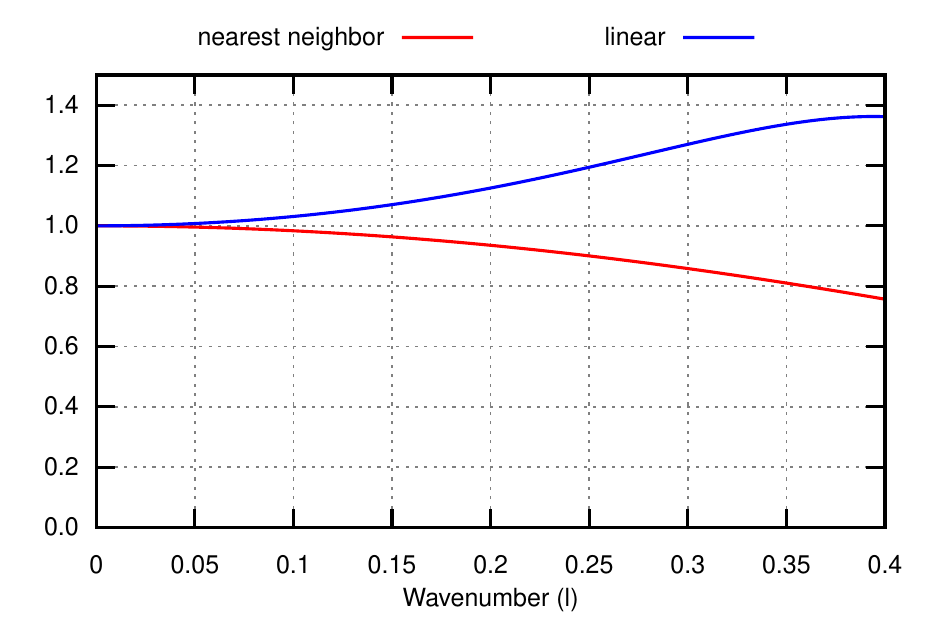}
	\caption{Comparison of the 1D pixel window for nearest neighbor
	mapmaking (red, $\text{sinc}(l)$) and linear mapmaking (blue). The 2D pixel window
	is the outer product of the 1D one along each axis. The pixel
	windows model the response of the Fourier coefficients to
	binned (unweighted) mapmaking. Square it to get the effect on
	the power spectrum.}
	\vspace{2mm}
	\label{fig:linwin1d}
\end{figure}

There will be no subpixel errors in the limit of infinitely small pixels,
so it's tempting to simply reduce the pixel size to solve the problem. This
does work, but since subpixel errors are proportional to $|\nabla s \cdot \vec \Delta|$,
where $s$ is the true, smooth signal on the sky and $\vec \Delta = [\Delta_x,\Delta_y]$
is the pixel shape, the improvement is only first order in the pixel side length.
A much more feasible solution is to instead improve the subpixel handling in the
pointing matrix. Going from nearest neighbor to bilinear interpolation is enough
to practically eliminate subpixel errors without reducing pixel size. We can
implement this in the toy example by defining
\begin{lstlisting}
pix_left = np.floor(pix).astype(int)
ry, rx  = pix-pix_left
iy1,ix1 = pix_left % N_side
iy2,ix2 = (pix_left+1)% N_side
weights = np.concatenate([
		(1-ry)*(1-rx), (1-ry)*rx,
		ry    *(1-rx), ry    *rx])
samps   = np.tile(np.arange(N_samp),4)
inds    = np.concatenate([
	iy1*N_side+ix1, iy1*N_side+ix2,
	iy2*N_side+ix1, iy2*N_side+ix2])
P_lin = scipy.sparse.csr_array((weights,(samps,inds)),shape=(N_samp,N_pix))
\end{lstlisting}
Bilinear mapmaking results in a different pixel window than
the standard \lstinline{pixwin_nn = sinc(ly)[:,None]*sinc(lx)[None,:]}
of nearest neighbor mapmaking. We derive this in appendix~\ref{sec:linwin},
and the result is \lstinline{pixwin_lin = linwin1d(ly)[:,None]*linwin1d(lx)[None,:]},
\lstinline{linwin1d(l) = sinc(l)**2/(1/3*(2-cos(2*pi*l)))}. This is plotted in
figure~\ref{fig:linwin1d}.

Since each sample in bilinear mapmaking gets contributions from the four closest pixels,
the pointing matrix is about four times slower than the standard nearest neighbor
version. However, as shown in figures~\ref{fig:2d-bias} and \ref{fig:2d-noise}
this cost is worth it, with bilinear maximum-likelihood mapmaking being bias-free
and even lower noise than standard ML. The bias is also greatly reduced for
destriping, but not eliminated.

Note that bilinear mapmaking (and in general any mapmaking method where each sample
affects multiple pixels) breaks the diagonality of $P^TP$ which methods like binning,
filter+bin and destriping rely on for their speed. While we have implemented bilinear
variants of them here in order to investigate the effect on the bias, bilinear filter+bin
and bilinear destriping are too slow to be useful in practice. This is in contrast to
bilinear maximum-likelihood mapmaking, which only becomes a few times slower.

\section{Gain errors}
Surprisingly, gain errors can also lead to a scale-dependent power loss.
This happens when the noise model correlates observations with inconsistent
gain, for example multipe miscalibrated detectors with correlated noise.
We will illustrate this using a minimal 1D toy example
with two correlated detectors, as well as a more realistic 2D example.

\subsection{1D toy example}
Consider two miscalibrated detectors scanning across a line of pixels. Both detectors
point in the same direction, and take one sample per pixel. We can model
this as
\begin{align}
	d &= G(Pm + n) = GPm + Gn & \text{(actual)} \label{eq:gain-toy1d}
\end{align}
where $d$ is the vector of samples, $m$ is the true signal in the pixels,
$n$ is Gaussian noise with covariance $N$, $P$ is the pointing matrix
and $G$ is the gain error. Written out in terms of detectors and samples,
we can express this as
\begin{align}
	d_{di} &= G_d P_d m_i + G_d n_{di}
\end{align}
with $d$ and $i$ being the detector and sample index respectively,
$P_d = [1, 1]_d = 1$ and $G_d = [g_1,g_2]_d$. This simply expresses
that both detectors see the same signal at the
same time. Let us assume that the noise covariance is diagonal in Fourier
space
\begin{align}
	N_{d_1d_2ff'} &= A_f C_{fd_1d_2} \delta_{ff'} \\
	C_{fd_1d_2} &= \begin{bmatrix}
	1 & \alpha_f \\
	\alpha_f & 1 \end{bmatrix}
\end{align}
where $f$ and $f'$ are frequency indices, and $A_f$ is the noise power spectrum.
This gives us the inverse noise model
\begin{align}
	N_{d_1d_2ff'}^{-1} &= A_f^{-1} C_{fd_1d_2}^{-1} \delta_{ff'} \label{eq:gain-toy1d-ninv}\\
	C_{fd_1d_2}^{-1} &= (1-\alpha_f^2)^{-1}\begin{bmatrix}
	1 & -\alpha_f \\
	-\alpha_f & 1 \end{bmatrix}
\end{align}
The maximum-likelihood solution of equation~\ref{eq:gain-toy1d} is
\begin{align}
	\hat m &= (P^TGG^{-1}N^{-1}G^{-1}GP)^{-1}P^TGG^{-1}N^{-1}G^{-1}d \notag \\
	&= (P^TN^{-1}P)^{-1}P^TN^{-1}(Pm + n) \notag \\
	&= m+(P^TN^{-1}P)^{-1}P^TN^{-1}n
\end{align}
which has expectation value $\langle \hat m \rangle = m$, which is unbiased.
However, given that $G$ is supposed to be a gain \emph{error}, we are presumably
not aware of it, and so our data model will instead be
\begin{align}
	d &= Pm + Gn & \text{(model)}
\end{align}
The reason why $G$ is still present in front of $n$ in the model is that the noise properties
are almost always measured from the data, so our noise model would automatically absorb $G$.
For example, if a detector were to be calibrated too high, it would also appear noisier, and would therefore be downweighted more under inverse variance weigting.
Given this model, our maximum-likelihood solution for $m$ is
\begin{align}
	\hat m &= (P^TG^{-1}N^{-1}G^{-1}P)^{-1}P^TG^{-1}N^{-1}G^{-1}d
\end{align}
Inserting the actual data behavior from equation~\ref{eq:gain-toy1d}, we get
\begin{align}
	\langle\hat m\rangle &= (P^TG^{-1}N^{-1}G^{-1}P)^{-1}P^TG^{-1}N^{-1}G^{-1}GPm \notag \\
	&= (P^TG^{-1}N^{-1}G^{-1}P)^{-1}P^TG^{-1}N^{-1}Pm
\end{align}
Since $N$ is diagonal in Fourier space and both $G$ and $P$ are constant in time,
the whole equation becomes Fourier-diagonal
\begin{align}
	\langle\hat m_f\rangle &= (P^TG^{-1}N_f^{-1}G^{-1}P)^{-1}P^TG^{-1}N_f^{-1}Pm_f
\end{align}
Inserting eq.~\ref{eq:gain-toy1d-ninv}, this simplifies to ($A_f$ cancels)
\begin{align}
	\langle\hat m_f\rangle &= g_1g_2\frac{g_1+g_2}{g_1^2+g_2^2} \frac{1-\alpha_f}{1-\frac{2g_1g_2}{g_1^2+g_2^2}\alpha_f} m_f
\end{align}
So the effect of the gain miscalibration is an overall scaling of the result
as one would expect, $g_1g_2(g_1+g_2)/(g_1^2+g_2^2)$, times a transfer function
\begin{align}
	\text{TF}(f) &= \frac{1-\alpha_f}{1-\frac{2g_1g_2}{g_1^2+g_2^2}\alpha_f}
\end{align}
This reduces to 1 both when the detectors are uncorrelated ($\alpha_f = 0$)
and when there are no relative gain errors ($g_1=g_2$).

\begin{figure}[ht]
	\centering
	\includegraphics[width=8cm]{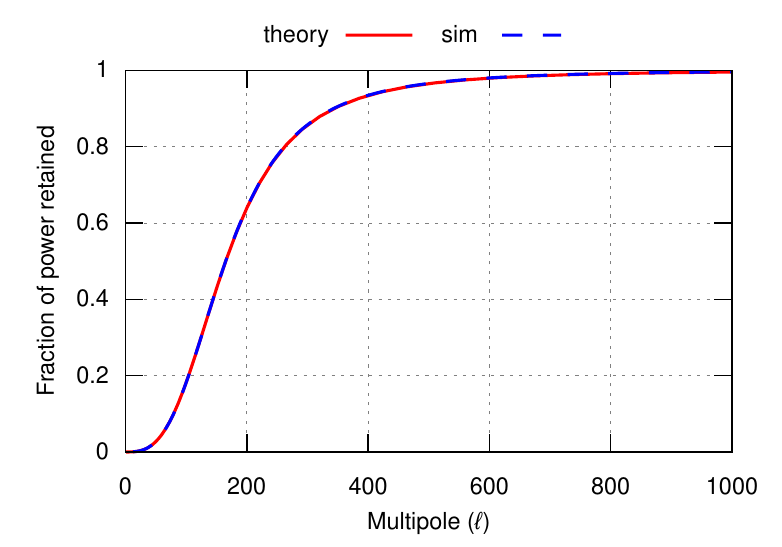}
	\caption{Signal loss for a 1D toy example with noise properties similar to
	a ground-based CMB telescope average weather (per-detector $\ell_\text{knee}=1000$)
	and a 10\% gain error. The solid red curve is based on equation~\ref{eq:gain-tf-1d}.
	The dashed blue curve is a simulation (\texttt{gain/gain\_model\_error\_toy\_1d\_sim.py}).
	Unlike the other examples, the horizontal axis
	uses multipoles instead of Nyquist units to make it easier to compare to
	CMB observations.}
	\label{fig:gain-tf-1d}
\end{figure}

To motivate a frequency-dependent correlation coefficient, let's consider the
common case of atmospheric 1/f-noise plus white instrumental noise. Our detectors
are co-pointing, meaning they see the same atmosphere, so this nosie component is
100\% correlated between detectors, but we assume their instrumental noise to be
uncorrelated.
\begin{align}
	N_f &= \begin{bmatrix}1 & 1 \\ 1 & 1\end{bmatrix} (f/f_\text{knee})^{-3} +
		\begin{bmatrix}1 & 0 \\ 0 & 1\end{bmatrix} =
			\begin{bmatrix}
				1 & \alpha_f \\ \alpha_f & 1
			\end{bmatrix} A_f \notag \\
	\alpha_f &= 1-A_f^{-1} \quad,\quad
	A_f = 1 + (f/f_\text{knee})^{-3} \quad \Rightarrow \notag \\
	\text{TF}(f) &= \frac{A_f^{-1}}{1-\frac{2g_1g_2}{g_1^2+g_2^2}(1-A_f^{-1})}
	\label{eq:gain-tf-1d}
\end{align}
The result is a frequency-dependent correlation coefficient that changes smoothly
from 100\% corrleated for atmosphere-dominated scales (low frequencies) to 0\% correlated
for detector-noise dominated scales (high frequencies). The transfer function for this
is plotted in figure~\ref{fig:gain-tf-1d} for the case of a 10\% gain error and
average weather (tough note that this 1D example does not capture the benefits of
crosslinking).

\subsection{2D toy example}
We have also demonstrated the effect of gain errors on a somewhat more realistic
2D simulation, where an array of four detectors scans first horizontally and then vertically
across a grid of $200\times200$ pixels, with one sample per pixel and each sample
hitting the pixel center to avoid subpixel errors.

The detectors are offset by
(0,0), (0,1), (1,0) and (1,1) pixels from the first so that the array instantaneously
observes more than just a single pixel, but still only a fraction of the whole image, much like
a real detector array would. Each detector has a (constant) gain error drawn
independently from a normal distribution with a standard deviation of 0.1, and both the
data and noise model reflect this.

The noise model consists of white noise plus a
common mode with a $1/\ell$ spectrum with a spectral index of $-3.5$ and $\ell_\text{knee} = 0.125$
(corresponding to 1/50th of the image side length). This common mode makes the noise model
strongly correlated on large scales, like the atmosphere does for real ground-based observations.
The bias requires inconsistent observations to be correlated, so the common mode is crucial for
this demonstration.

We solve for binned, maximum-likelihood and filter+bin
maps using a data model that is unaware of the gain errors (as one would be in reality,
or they wouldn't be gain errors). This is implemented in the program \verb|gain_mode_error_2d_toy.py|
(see section~\ref{sec:code}).

\begin{figure}
	\centering
	\includegraphics[width=8.5cm]{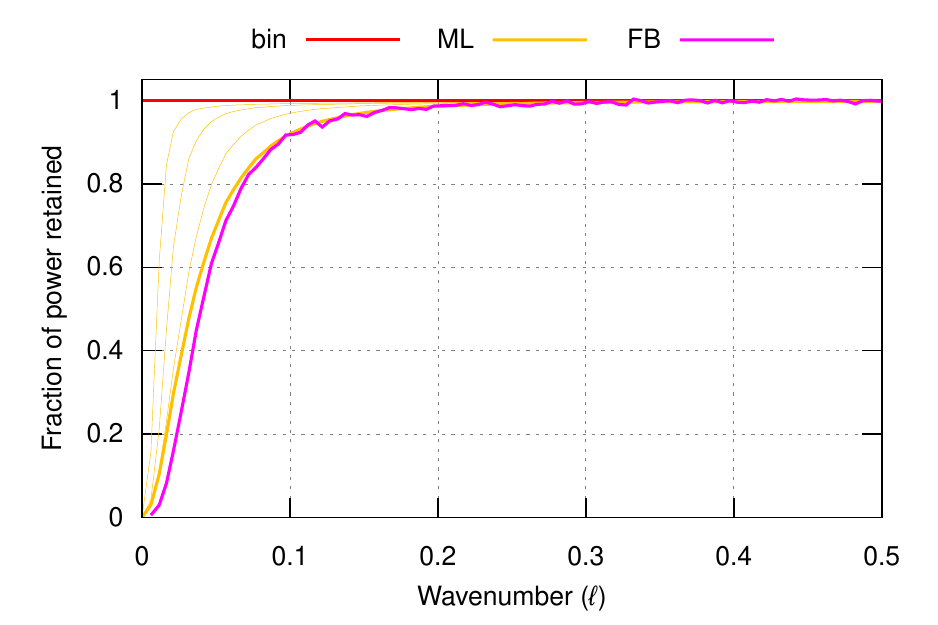}
	\caption{Power loss for a 2D toy example with multiple detectors
	with inconsistent but constant gain errors. Plain unweighted binning
	(red) is unbiased, but both maximum-likelihood (yellow) and filter+bin
	(pink) lose much of the signal at large scales (low wavenumber).
	The light blue curves are maximum-likelihood with 10/100/1000 times
	as much white noise in the noise model, reducing the overall correlatedness of the noise.
	This results in less bias, but also much more noise as shown in figure~\ref{fig:gain-noise-2d}.}
	\label{fig:gain-tf-2d}
\end{figure}

\begin{figure}
	\centering
	\includegraphics[width=8.5cm]{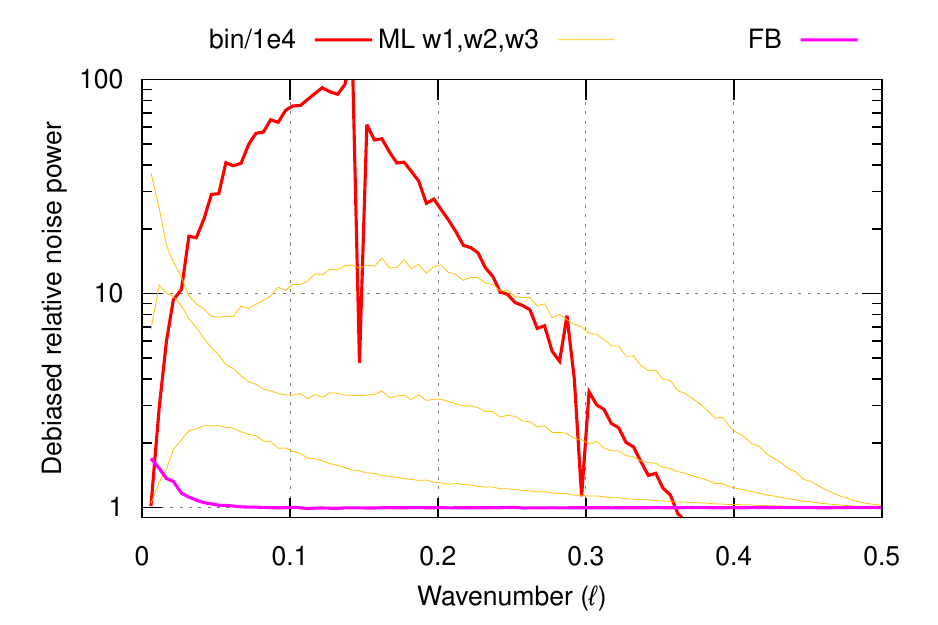}
	\caption{How noisy different mapmaking methods are compared to maximum-likelihood
	for the 2D gain error toy example. Each graph is the ratio of the transfer-function-deconvolved
	noise spectrum for binned mapmaking (red), whitened maximum-likelihood (yellow)
	or filter+bin (pink) divided by the corresponding spectrum for standard maximum-likelihood.
	The whitened curves correspond to a noise model with 10/100/1000 times exaggerated
	white noise floor for the bottom/middle/top light blue curve. Both these and binned mapmaking
	are too costly to be realistic solutions to the power loss problem. Note the logarithmic
	y axis and that the binned curve has been divided by $10^4$ to fit in the graph.}
	\label{fig:gain-noise-2d}
	\vspace{3mm}
\end{figure}

As expected from the 1D toy example, the gain inconsistency results
in a large loss of power on large scales, as shown in figure~\ref{fig:gain-tf-2d}.
Filter+bin mapmaking is no less vulnerable to this bias than maximum-likelihood.
Figure~\ref{fig:gain-noise-2d} shows the corresponding relative noise spectra.
And as we saw for subpixel errors, attempting to mitigate the bias by
artificially reducing the correlations in the noise model is much too
costly noise-wise to be a practical mitigation strategy.\footnote{In filter+bin
mapmaking the equivalent to reducing the correlations in the noise model would be
to filter less.}

Detectors with strongly correlated noise on large scales and O(10\%) gain
error inconsistency are not unrealistic for a ground-based CMB telescope,
so it is likely that many ground-based telescopes suffer from this bias to
some extent.

\subsection{Time-dependent gain errors are much less serious}
\label{sec:toy-gain1d-time}
So far we have only considered cases where detectors disagree on the gain,
but gain errors can also be time-dependent. Can these also cause large-scale
power loss? As we shall see the answer is ``yes, but only in unrealistic
circumstances''.

Consider a single detector that makes a series of passes over a line of pixels,
such that

\begin{align}
	d_{ix} &= g_{ix} m_x + g_{ix} n_{ix} \quad \text{(actual)}
\end{align}
where $d_{ix}$ is the measurement in pixel/sample $x$ of pass $i$, $m_x$ is the
true signal in pixel $x$, $g_{ix}$ is the unmodeled gain and $n_{ix}$ is the
noise. We assume that the gain error is ignored in the data model but captured
by the noise model:
\begin{align}
	d_{ix} &= m_x + g_{ix} n_{ix} \quad \text{(model)}
\end{align}
This has maximum-likelihood solution
\begin{align}
	\left(\sum_i g_{ix}^{-1} N_{xy}^{-1} g_{iy}^{-1}\right) \hat m_y &= \sum_i g_{ix}^{-1}N_{xy}^{-1} (m_y + n_{iy})
\end{align}
We further assume
\begin{enumerate}
	\item Both the noise and gain errors are independent between each pass
	\item The inverse gain perturbation $\gamma_{ix} = g_{ix}^{-1} - 1$
		has stationary covariance $\Gamma$, which is therefore diagonal
		in Fourier space, with diagonal $\widetilde{\Gamma}$ (its power spectrum).
	\item Similarly the covariance is Fourier-diagonal with power spectrum
		$\widetilde{N}$
\end{enumerate}
Under these assumptions it can be shown (see appendix~\ref{sec:toy-gain1d-time-deriv})
that when the number of passes is much
greater than one the Fourier transform of the maximum likelihood solution becomes
\begin{align}
	\hat{\widetilde{m}}_f &= \frac{\widetilde{N}^{-1}_f}{\widetilde{N}^{-1}_f + S^{-2}(\widetilde{\Gamma} \ast \widetilde{N}^{-1})_f} \widetilde{m}_f
\end{align}
where $\ast$ is the convolution operator and $S$ is the number of pixels, and we've assumed the standard discrete Fourier transform normalization. We see that in the limit of uncorrelated
gain errors, $\widetilde{\Gamma}_f = \text{const}$, the bias becomes $1/(1+\text{const}\cdot\widetilde{N})$.
For atmospheric noise this gives a strong suppression of the signal on large scales, just
like we saw for detector errors. On the other limit where the gain error changes slowly
and is approximately constant over a noise correlation length, we have $\widetilde{\Gamma} \approx \text{const}\cdot \delta \Rightarrow \widetilde{\Gamma} \ast \widetilde{N}^{-1} \approx \text{const} \cdot \widetilde{N}^{-1}$,
and so the bias just becomes a constant scaling. So to get scale-dependent power loss,
the gain error needs to vary on time-scales over which the noise is strongly correlated.

However, all of this rests on the assumption that the gain errors are
captured by the noise model. For the previous case of constant per-detector
gain errors this is realistic - after all detectors typically don't all have the
same noise level, so they need to be weighted differently, and the noise model
ends up absorbing the gain error when being measured from the data itself.
The situation is different
from time-dependent gains. As we saw above, the noise model must be modulated
by gain errors that change quickly relative to the atmosphere if we are to
get scale-dependent bias, and for a typical ground-based CMB telescope this
works out to be $\sim$ second time-scales. It is very uncommon to build noise models
that track the noise properties with a time resolution this high.

In contrast
to detector-dependent gain errors, we therefore
do not expect time-dependent gain errors to be a significant source for large-scale
power loss.

\section{Easily missed in simulations}
What makes model error bias especially insidious is that
it is completely invisible to any simulation that does not explicitly include
that particular type of model error. For example, a simulation where the CMB
is read off from a simulated map using the same pointing matrix as is used
in the mapmaking itself would be blind to subpixel errors. And simulations that
don't inject gain errors into both the data and noise model will be blind to
power loss from gain errors. Given the many
possible ways the real data might deviate from one's model of it, it is hard
to be sure that one has included all the relevant types of model error in the simulations.
It is therefore very easy to trick oneself into believing that one has an
unbiased analysis pipeline while there are in fact $\mathcal{O}(1)$ biases remaining.

\section{Detection strategies}
The unintuitive large-scale effects of model errors originate from the interaction
between model errors and non-local weighting/filtering. A noise model (or filter) with
a large dynamic range, such as one capturing the huge ratio between the long-wavelength
and short-wavelength noise power typical for ground-based CMB telescopes, is therefore
much more vulnerable to large-scale power loss than one appropriate for
space-based telescopes which have almost flat noise power spectra. This suggests
the following tests for large-scale model error bias:
\begin{enumerate}
	\item Compare power spectra with those from a space-based telescope.
		This is reliable but comes at the cost of being able to make an independent measurement.
	\item Split the data into subsets with different ratios of correlated noise to white noise
		and check their consistency. This could for example be a split by the
		level of precipitable water-vapor (PWV) in the atmosphere. High-PWV-data
		would be expected to have a higher dynamic range and therefore be more vulnerable.
	\item Map the same data both using the standard noise model/filter and a
		less optimal one an artificially reduced lower dynamic range, e.g. one that
		underestimates the amount of correlated noise or ignores correlations between
		detectors. The latter should result in less a biased but noisier map, as we saw
		in figures~\ref{fig:2d-bias} and \ref{fig:2d-noise}.
		If the maps are consistent, then large-scale model error bias is probably not an issue.
		Alternatively, one can check for consistency when varying the pixel size. We haven't
		tested the effectiveness of this method, but model error bias should be proportional
		to the pixel size with nearest neighbor mapmaking.
\end{enumerate}

\section{Who is at risk?}
\begin{figure}[ht]
	\centering
	\includegraphics[width=\columnwidth]{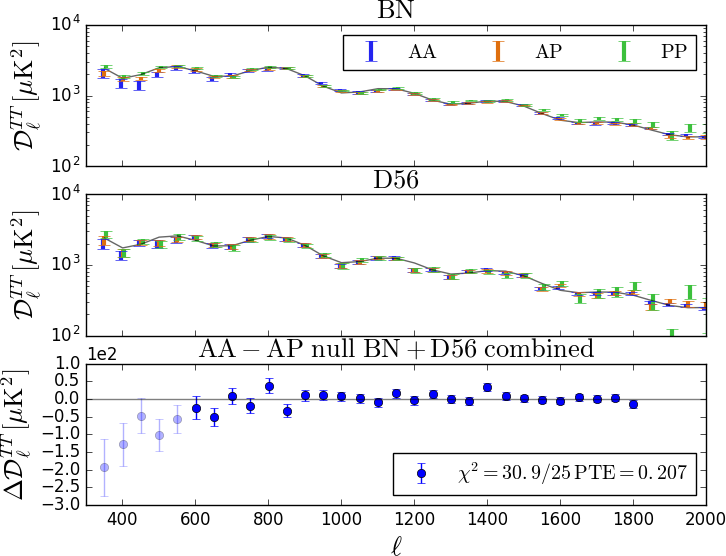}
	\caption{Large-scale power loss in ACT DR4. The ACT TT power spectrum (blue)
	and the ACT-Planck cross-spectrum (orange) are both low at $\ell \lesssim 800$
	compared to the Planck power spectrum (green) in the two ACT patches shown.
	This is especially clear in the third panel, which shows the difference between
	ACT and the ACT-Planck cross. Reproduced from \citet{choi/etal:2020}.}
	\label{fig:act-dr4-tf}
\end{figure}

While model error bias can be expected to always be present at a
low level, it only becomes important when dealing with very large amounts
of correlated noise (e.g. much more noise correlations on some scales than others,
see equation~\ref{eq:gain-tf-1d}). Therefore, the telescopes at risk are
those that fulfill the following criteria.
\begin{enumerate}
	\item Observe from the ground.
	\item Care about total intensity, since the atmosphere is mostly unpolarized.
	\item Observe at frequencies with high water vapor emission, since
		water vapor is the main source of clumpiness in the atmospheric emission.
		At CMB-relevant frequencies, the lower the frequency the safer.
		Frequencies $<60$ GHz should be relatively safe while $>200$ GHz are
		particularly vulnerable.
	\item Have sensitive detectors, since lower white noise means a larger
		ratio between this noise and the correlated noise from the atmosphere.
	\item Have a large number of detectors with a low angular separation on
		the sky, since this increases their correlatedness.
\end{enumerate}
There aren't that many telescopes that fulfill these criteria, as ground-based
telescopes have focused mostly on polarization or lower frequencies. After reviewing
\href{https://lambda.gsfc.nasa.gov/product/expt}{LAMBDA's list of CMB experiments}
we found four $>60$ GHz ground-based CMB telescope projects started after year 2000\footnote{
	Ground-based projects older than this are too insensitive for this bias to be visible even if
	it were to be present.}
with published TT spectra for $\ell < 2000$: ACBAR, ACT, BICEP and SPT. Of these,
ACBAR, ACT \citep{choi/etal:2020} and SPT-3G \citep{spt3g-ps-2022} cut their low-$\ell$ data, indicating problems recovering
these scales. Meanwhile BICEP \citep{barkats/etal/2013, bicep-keck-ps-2018} and SPTpol \citep{sptpol-ps-2018} publish TT spectra down to $\ell < 100$ with
no signs of problems. The only case where we could see clear evidence for large-scale
power-loss is in the multipoles just above the multipole cutoff in ACT, as seen in
figure~\ref{fig:act-dr4-tf} (reproduced from \citet{choi/etal:2020}).

So to summarize, why hasn't this been noticed before? Ground-based CMB surveys
have mostly focused on polarization, where this effect is orders of magnitude
smaller. Those that do publish total intensity power spectra
have only reached the high enough sensitivity and detector counts for it to
become noticable in the last decade. Given how frequently ground-based CMB
surveys cut $\ell \lesssim 500-1000$ it is tempting to speculate that model error
bias may have caused null tests to fail at low $\ell$ without being
identified as the culprit.

\section{Source code}
\label{sec:code}
The source code and data files behind these examples is available at
\url{https://github.com/amaurea/model_error2}. The 1D and 2D subpixel
simulations are available in \verb|subpix/model_error_toy.py| and \verb|subpix/model_error_toy.py|
respectively, while the gain error example is in
\verb|gain/gain_model_error_toy_2d.py|.

\section{Conclusion}

\begin{enumerate}
	\item \textbf{All current CMB mapmaking methods are vulnerable to model error bias},
		including maximum-likelihood mapmaking, destriping and filter+bin mapmaking.
	\item \textbf{Subpixel errors} and \textbf{gain errors} are common examples of
		model errors, with subpixel errors being almost universal.
		Subpixel bias is caused by the likelihood disfavoring the sharp edges implied by
		the pixelization, causing it to discard low-S/N signal. Gain errors enter into inverse
		variance weighting squared, leading to a noise bias term that dominates strongly downweighted/filtered scales.
		Other types of model error can also be important.
	\item Model error bias can manifest in unintuitive ways, with a common symptom
		being a \textbf{large ($\mathcal{O}(1)$) loss of long-wavelength power} in the maps.
	\item The size of the bias is proportional to the dynamic range of the filter/noise model.
		Hence it is \textbf{important for ground-based measurements of the unpolarized CMB}
		due to the presence of large-scale atmospheric noise there, but is \textbf{much less
		important for polarization measurements or space-based telescopes}.
	\item \textbf{Simulations are blind to these biases unless specifically designed to
		target them}. There is a large risk of ending up thinking one has an unbiased
		pipeline despite there being large bias in the actual CMB maps.
	\item An effective way of testing for this bias is to \textbf{also map the data using a
		(much) lower dynamic range noise model/filter}\footnote{Or much longer baselines for destriping}
		and checking if this leads to consistent power spectra.
\end{enumerate}

While it's unclear how large a role model-error bias has played in ground-based CMB results published
so far, it will be an increasingly important effect as sensitivity increases. We hope upcoming
projects like Simons Observatory and CMB-S4 will be on the lookout for large-scale power loss in
total intensity, and to be aware that simulations alone cannot prove that their analysis pipeline
is bias free.

\section*{Acknowledgements}
We would like to thank Jon Sievers for first making us aware
that subpixel errors weren't just limited to X'es around point
sources and other small-scale effects, and Ed Wollack and Lorenzo Piazzo for useful comments.
Yunyang Li pointed out an error in the original version of the 1D gain
toy example (now fixed).
This work was supported by a grant from the Simons Foundation (CCA 918271, PBL).

\bibliographystyle{act_titles}
\bibliography{refs}

\clearpage

\appendix

\section{Linear pixel window}
\label{sec:linwin}
\begin{figure}[h]
	\center
	\includegraphics[width=5cm]{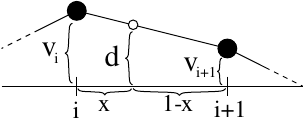}
	\caption{Definition of the quantities involved in linear interpolation.
	Given values $\{v_i\}$ sampled at locations $\{i\}$, we define the linear interpolation
	value $d$ at some location $i+x$ where $x\in[0,1)$ as $d = (1-x)v_i + x v_{i+1}$.}
	\label{fig:linear-defs}
\end{figure}

Let us start by considering the 1D case. Given a set of pixels $\{i\}$ with values $\{v_i\}$
we define the linear interpolation value $d$ at some pixel coordinate $i+x$ where $x\in[0,1)$ as
\begin{align}
	d = (1-x)v_i + x v_{i+1}
\end{align}
This is illustrated in figure~\ref{fig:linear-defs}.

\begin{figure}[h]
	\center
	\includegraphics[width=12cm]{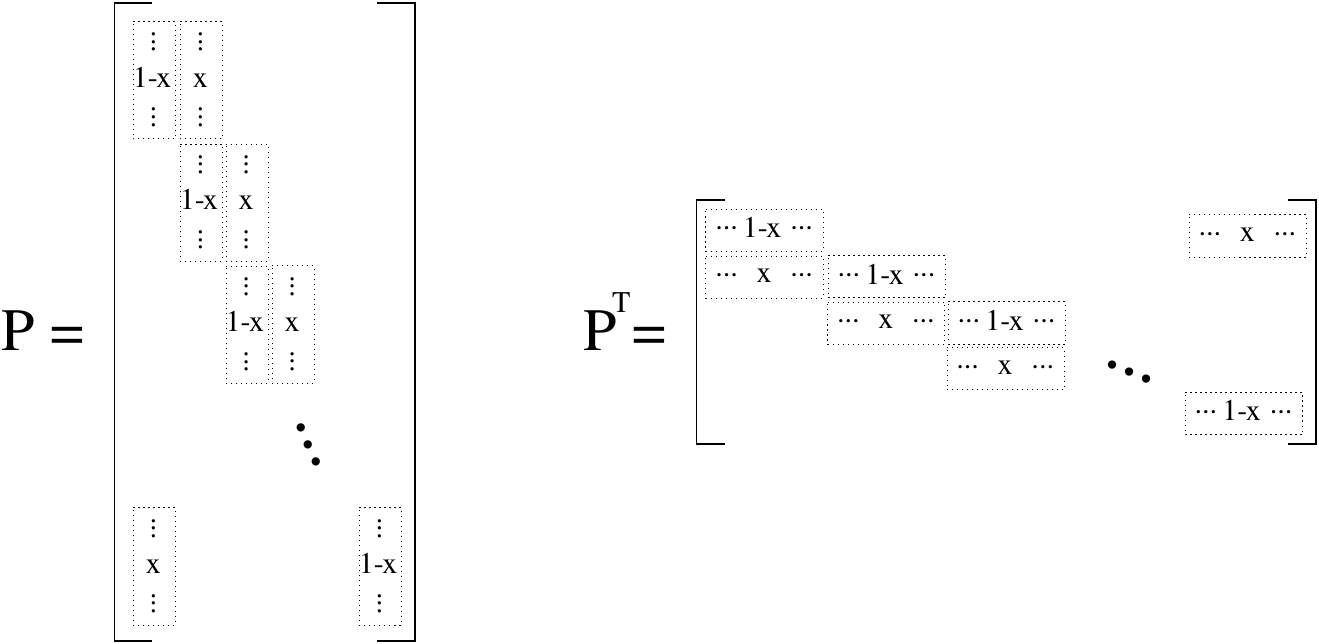}
	\caption{Structure of a linear-interpolation pointing matrix $P$ and its transpose $P^T$
	for the case of a set of samples making a single scan across a row of pixels in 1D, assuming
	wrapping boundary conditions. $x \in [0,1)$ is the subpixel coordinate of each sample.}
	\label{fig:linear-p}
\end{figure}

We can generalize this to a set of samples $d_a$ with
\begin{align}
	d_a &= (1-x_a) v_a^\text{left} + x_a v_a^\text{right}
\end{align}
where $v_a^\text{left}$ and $v_a^\text{right}$ are the values of the pixel to the left and right
of where sample $a$ points, and $x_a$ is its subpixel position. This linear system can be written
as the following matrix equation
\begin{align}
	d &= P v
\end{align}
$P$'s form will depend on the exact scanning pattern, but the final pixel window should only depend
on each pixel being hit equally much at all subpixel locations. For simplicity, we will assume a simple
scanning pattern that scans a single time across a row of pixels from left to right, with a constant
scanning speed and infinite sample rate. This results in the pointing matrix $P$ taking the form
shown in figure~\ref{fig:linear-p}.

Given the set of samples $d$, the least squares estimator for the
pixel values $v$ under uniform weighting is
\begin{align}
	\hat v &= (P^TP)^{-1}P^T d
\end{align}
The denominator $P^TP$ consists of two types of non-zero cases:
\begin{align}
	(P^TP)_{00} &=
		\begin{bmatrix}\cdots & 1-x & \cdots\end{bmatrix}
		\begin{bmatrix}\vdots\\1-x\\\vdots\end{bmatrix} +
		\begin{bmatrix}\cdots & x & \cdots\end{bmatrix}
		\begin{bmatrix}\vdots\\x\\\vdots\end{bmatrix}
			= S \int_0^1 (1-x)^2 dx + S \int_0^1 x^2 dx = \frac{2}{3}S \\
	(P^TP)_{01} &=
		\begin{bmatrix}\cdots & 1-x & \cdots\end{bmatrix}
		\begin{bmatrix}\vdots\\x\\\vdots\end{bmatrix}
			= S \int_0^1 (1-x)x dx = \frac{1}{6}S
\end{align}
Here $S$ is the number of samples per pixel, and appears due to the summing implied by $P^T$.
$S\rightarrow\infty$ in the integral limit we used here,
but it will end up cancelling. With this, we end up with the denominator taking the Toeplitz form
\begin{align}
	P^TP &= \frac{1}{6}N\begin{bmatrix}
		4 & 1 &   &   &  &  & 1 \\
		1 & 4 & 1 &   &  &  & \\
			& 1 & 4 & 1 &  &  & \\
			&   &   & \ddots & & & \\
		  &   &   &   & 1 & 4 & 1 \\
		1 &   &   &   &   & 1 & 4
	\end{bmatrix}
\end{align}
Toeplitz matrices are diagonal in Fourier space, so $(P^TP)^{-1}$ becomes a simple element-wise division
there. The Fourier-diagonal is given by the Fourier transform of the first row of the Toeplitz matrix.
Defining $A = P^TP$, it's Fourier diagonal as $\tilde a$ is given by
\begin{align}
	\tilde a(k) &= \frac{1}{N} \sum_{j=0}^{N-1} e^{\frac{-2\pi ikj}{N}} A_{j0}
	= \frac{S}{N} \frac{1}{6} \Big(4 + e^{\frac{2\pi ik}{N}} + e^{\frac{-2\pi ik}{N}} \Big)
	= \frac{S}{N} \frac{1}{3} \Big(2 + \cos\frac{2\pi k}{N}\Big)
\end{align}
where $N$ is the number of pixels, $i$ is the imaginary unit, $k$ is the zero-based index into $\tilde a$
and $j$is the zero-based row index of $A$.

\begin{figure}[ht]
	\centering
	\includegraphics[width=8cm]{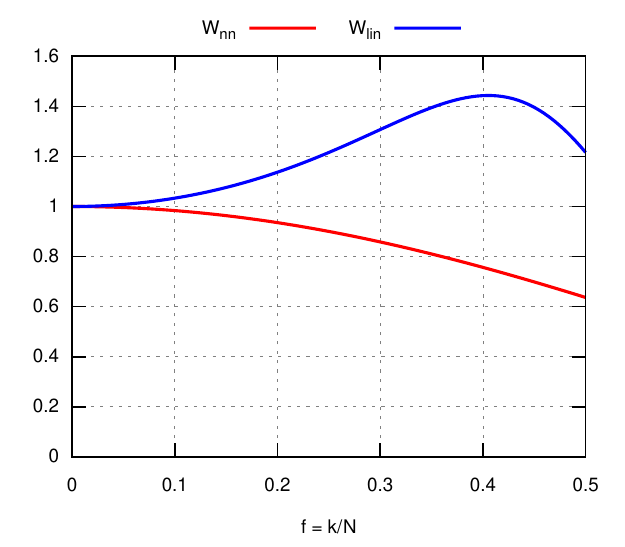}
	\caption{Comparison the 1D pixel windows for nearest neighbor mapmaking ($W_\text{nn}$)
	and linear mapmaking ($W_\text{nn}$), plotted up to the Nyquist frequency, which is $\frac12$ in
	these units.}
	\label{fig:pixwins}
\end{figure}

Turning to the numerator $b = P^Td$, we recognize that it is simply the convolution of $d$ with
the triangle function $T(x) = (1-|x|)S, x\in [-1,1]$, evaluated at each pixel center.\footnote{
This means that $T*d$ is the same function as $b$, just $S$ times more densely sampled. Hence,
it has the same Fourier representation up to $b$'s Nyquist frequency.}
Using the convolution theorem, we can therefore directly write down
\begin{align}
	\tilde b(k) &= \tilde T(k) \cdot \tilde d(k)
\end{align}
We're left with computing the Fourier-space representation of the triangle function $T$.
\begin{align}
	\tilde T(k) &= \frac{S}{N} \int_{-1}^1 e^\frac{-2\pi i k x}{N} (1 - |x|) dx \notag \\
	&= \frac{S}{N}\left[\int_{-1}^1 e^\frac{-2\pi i k x}{N} dx - \int_0^1 e^\frac{-2\pi i k x}{N} x dx - \int_{-1}^0 e^\frac{-2\pi k x}{N} (-x) dx\right] \notag \\
	&= \frac{S}{N}\left[ 2 \sinc\frac{2k}{N} -  2\sinc\frac{2k}{N} + 2\Big(\frac{2\pi k}{N}\Big)^2 \Big(1-\cos\frac{2\pi k}{N}\Big)\right] \notag \\
	&= \frac{S}{N}\Big(\sinc\frac{k}{N}\Big)^2
\end{align}
Combining the numerator and denominator we have
\begin{align}
	\hat {\tilde v} &= \frac{\tilde b(k)}{\tilde a(k)} \tilde d(k) = \frac{\big(\sinc\frac{k}{N}\big)^2}{\frac{1}{3}\big(2+\cos\frac{2\pi k}{N}\big)} \tilde d(k) = W_\text{lin}(k/N) \tilde d(k)
\end{align}
where the linear interpolation pixel window $W_\text{lin}(f)$ is defined as
\begin{align}
	W_\text{lin}(f) &= \frac{\sinc(f)^2}{\frac{1}{3}\big(2+\cos(2\pi f)\big)}
\end{align}
This pixel window is compared to the standard nearest neighbor pixel window $W_\text{nn} = \sinc(f)$
in figure~\ref{fig:pixwins}. Their behavior is quite different. While $W_\text{nn}$ falls monotonically
to around 0.6 at the Nyquiest frequency, $W_\text{lin}$ rises to about 1.4 at $f = 0.4$, and only then
starts falling.

The natural generalization of linear interpolation to two dimensions is bilinear interpolation,
which is the composition of a horizontal and vertical one-dimensional interpolation. Because this
operation is separable, so is the resulting pixel window. Hence
\begin{align}
	W_\text{lin}^\text{2D}(f_y,f_x) &= W_\text{lin}(f_y) W_\text{lin}(f_x)
\end{align}

\vspace{5mm}

\section{Time-dependent gain errors}
\label{sec:toy-gain1d-time-deriv}
Continuing from section~\ref{sec:toy-gain1d-time}, we had
\begin{align}
	\underbrace{\left(\sum_i g_{ix}^{-1} N_{xy}^{-1} g_{iy}^{-1}\right)}_{A_{xy}} \hat m_y &= \underbrace{\sum_i g_{ix}^{-1}N_{xy}^{-1} (m_y + n_{iy})}_{b_x}
\end{align}
Expanding $g_{ix}^{-1} = 1+\gamma_{ix}$ and assuming the detector makes $M$ passes over the pixels, we get
\begin{align}
	A_{xy} &= \sum_i^M(1+\gamma_{ix})N_{xy}^{-1}(1+\gamma_{iy}) \notag \\
	&= M\Big(N_{xy}^{-1} + \underbrace{\frac1M \sum_i^M \gamma_{ix}N_{xy}^{-1}}_\text{odd} + \underbrace{\frac1M \sum_i^M N_{xy}^{-1}\gamma_{iy}}_\text{odd} + \underbrace{\frac1M \sum_i^M \gamma_{ix}N_{xy}^{-1}\gamma_{iy}}_{\psi_{xy}}\Big)
\end{align}
In the limit $M\rightarrow\infty$ the odd powers of the random variable $\gamma$ disappear,
so we are left with computing $\psi$. Since both $N$ and $\Gamma$ are diagonal in Fourier
space, it makes sense to express $\psi$ in Fourier space. Denoting Fourier-space quantities with
a tilde, and defining the Fourier transform operators
\begin{align}
	F_{kx} &= \exp(-2\pi \sqrt{-1} k x / S) &
	F_{xk}^{-1} &= \frac1N \exp(2\pi \sqrt{-1} k x / S)
\end{align}
where $S$ is the number of pixels, we get
\begin{align}
	\widetilde{\psi}_{jk} &= \frac1M \sum_i^M \sum_{xy} F_{jx}(F^{-1}\widetilde\gamma)_x (F^{-1}\widetilde{N}F^{-1\dag})^{-1}_{xy}(F^{-1}\widetilde\gamma)_y^\dag F_{yk}^\dag \notag \\
	&= \frac1M \sum_i^M \sum_{xy} F_{jx}(F^{-1}\widetilde\gamma)_x (F^\dag \widetilde{N}^{-1}F)_{xy}(F^{-1}\widetilde\gamma)_y^\dag F_{yk}^\dag \notag \\
	&= \frac1M \sum_i^M \sum_{xy\alpha\beta\gamma} \frac{1}{N^2}\exp\Big(\frac{2\pi\sqrt{-1}}{S}(\underbrace{-jx+\alpha x+\beta x}_{N\delta_{\beta,j-\alpha}} \underbrace{-\beta y -\gamma y + ky}_{N\delta_{\gamma,k-\beta}})\Big) \widetilde\gamma_\alpha \widetilde{N}_{\beta}^{-1} \widetilde\gamma_\gamma^\dag\Big) \notag \\
	&= \sum_\alpha \underbrace{\Big(\frac1M \sum_i^M \widetilde\gamma_\alpha \widetilde\gamma_{k-j+\alpha}\Big)}_{\widetilde\Gamma_{\alpha,\alpha+k-j} = \widetilde\Gamma_\alpha \delta_{k,j}} \widetilde{N}_{j-\alpha}^{-1} \notag \\
	&= \sum_\alpha \widetilde\Gamma_\alpha \widetilde{N}_{j-\alpha}^{-1} \delta_{jk} \notag \\
	&= \Big(\widetilde\Gamma \ast \widetilde{N}^{-1}\Big)_j \delta_{jk}
\end{align}
So $\psi$ is Fourier diagonal just like $\Gamma$ and $N^{-1}$, and its diagonal is
the convolution of their diagonals. With this we can complete the expression for $A$.
\begin{align}
	\widetilde{A} &= F A F^\dag
	= F\Big((F^{-1}\widetilde{N}F^{-1\dag})^{-1} + F^{-1}\widetilde\psi F^{-1\dag}\Big)F^\dag
	= MS^2\Big(\widetilde{N}^{-1} + S^{-2} \widetilde\Gamma \ast \widetilde{N}^{-1}\Big)
\end{align}
Moving on to the right-hand-side of the equation for $\hat m$, we have
\begin{align}
	b_x &= \sum_i^M(1+\gamma_{ix})N_{xy}^{-1}(m_y + n_{iy}) \notag \\
	&= M\Big(N_{xy}^{-1}m_y + \underbrace{\frac1M\sum_i^M\gamma_{ix}N_{xy}^{-1}n_{iy}}_\text{odd} + \underbrace{\frac1M\sum_i^M \gamma_{ix}N_{xy}^{-1} m_y}_\text{odd} + \underbrace{\frac1M\sum_i^M\gamma_{ix}N_{xy}^{-1}n_{iy}}_\text{$\gamma$ and $n$ independent}\Big)
\end{align}
All but the first term fall out if we assume that the fluctuations in $\gamma$ and $n$
are independent. Going to Fourier space, we are left with
\begin{align}
	\widetilde b &= Fb = MF\Big(F^{-1}\widetilde{N}F^{-1\dag}\Big)^{-1} F^{-1}\widetilde{m}
	= M\underbrace{FF^\dag}_{S}\widetilde{N}^{-1}\underbrace{FF^{-1}}_{I}\widetilde{m}
	= MS\widetilde{N}^{-1}\widetilde{m}
\end{align}
With this the full Fourier-space equation for $\hat{\widetilde{m}}$ becomes:
\begin{align}
	A\hat m &= b \Leftrightarrow \notag
	F^{-1}\widetilde{A}\underbrace{F^{-1\dag}F^{-1}}_{S^{-1}}\hat{\widetilde{m}} = F^{-1}\widetilde{b} \Leftrightarrow \notag
	\hat{\widetilde{m}} = \frac{\widetilde{N}^{-1}}{\widetilde{N}^{-1} + S^{-2}\widetilde{\Gamma}\ast\widetilde{N}^{-1}} \widetilde{m}
\end{align}
The factor $S^{-2}$ is an artifact of the Fourier normalization used, and would
differ for other Fourier unit choices.

\end{document}